\documentclass[nonacm]{acmart}

\AtBeginDocument{%
  }

\usepackage{rotating} % add this to your preamble
\usepackage{tabularx}
\usepackage{longtable}
\usepackage{subcaption}
\providecommand{\noopsort}[1]{}

\begin{document}

\title{Household Plastic Recycling: Empirical Insights and Design Explorations}

\author{Ashley Colley}
\orcid{0009-0008-5967-2828}
\affiliation{%
  \institution{University of Lapland}
  \city{Rovaniemi}
  \country{Finland}
}
\email{ashley.colley@ulapland.fi}

\author{Emma Kirjavainen}
\orcid{0000-0002-5181-6865}
\affiliation{%
  \institution{University of Lapland}
  \city{Rovaniemi}
  \postcode{96300}
  \country{Finland}}
\email{emma.kirjavainen@ulapland.fi}

\author{Sari Tapio}
\orcid{}
\affiliation{%
  \institution{University of Lapland}
  \city{Rovaniemi}
  \postcode{96300}
  \country{Finland}}
\email{sari.tapio@ulapland.fi}

\author{Jonna Häkkilä}
\orcid{0000-0003-2172-6233}
\affiliation{%
  \institution{University of Lapland}
  \city{Rovaniemi}
  \country{Finland}
}
\email{jonna.hakkila@ulapland.fi}

\renewcommand{\shortauthors}{Colley et al.}

\begin{abstract}
This article examines household plastic recycling in Finland through two qualitative studies and four design concepts. Study 1 reports short interviews with residents about how they store, sort, and dispose of plastic packaging in their homes. The findings highlight recurring frictions: limited space, improvised storage, uncertainty about correct sorting, and difficulties with bulky or dirty items. Study 2 focuses on laundry detergent packaging as a common source of large plastic containers. Participants' purchase decisions prioritised price and cleaning performance, while expressing concern for environmental impact and confusion about materials, rinsing, and recyclability.

Building on these insights, four student groups designed interactive recycling concepts that combine physical bins or bags with mobile applications. The concepts explore modular storage, sensing and compaction, playful feedback, and reward schemes to support domestic recycling routines. Together, the studies and concepts point to design opportunities at the intersection of packaging, home infrastructure, and digital services, while also raising questions about feasibility, privacy, and the cost of new devices.
\end{abstract}

\maketitle

\section{Introduction}

\begin{figure}
    \centering
\includegraphics[width=0.75\linewidth]{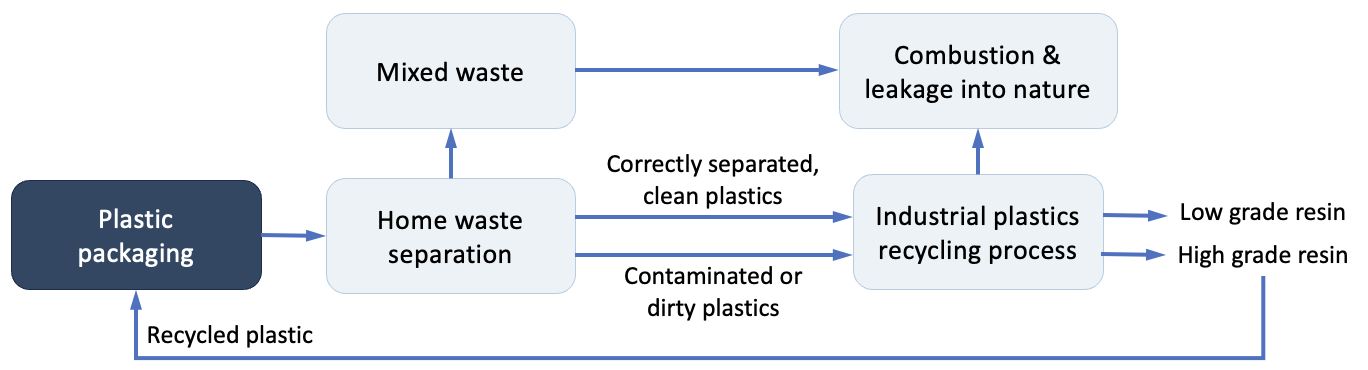}
    \caption{Overview of the plastic packaging recycling process, highlighting the central position of recycling in the home environment}
    \label{fig:process_diag}
\end{figure}

Plastic packaging supports food safety, transport, and storage, yet its long life and accumulation in ecosystems raise concerns \cite{nasrollahi2020plastic}. Public debate has grown as production continues to rely on virgin materials and disposal practices fail to keep pace with rising waste volumes \cite{gabbatiss2018world}. These tensions place household recycling practices under scrutiny, since homes form the first step in sorting and preparing packaging for reuse or material recovery (Figure \ref{fig:process_diag}). Prior studies show that clear instructions, convenient routines, and low effort strongly influence waste sorting outcomes in everyday life \cite{kokkonen2020kierratyksen,oluwadipe2022critical}. 

Finland faces this challenge directly. The current recycling rate for plastic packaging stands at roughly 29\%, and national targets require a large increase over the next years. In 2023 the European Commission issued an early warning that Finland may miss the upcoming targets \cite{EU2023}. This highlights the need for solutions that do more than raise collection volumes. Packaging design, clear labelling, suitability for machine sorting, and how packaging fits into household routines all influence whether material cycles perform as intended. Misplaced items, poorly cleaned containers, or plastic types that are hard to recognise may reduce recycling quality or steer recoverable material into mixed waste. At the same time, excessive cleaning with hot water or detergents may offset environmental gains, adding a further layer of tension to everyday decisions.

This article examines these issues through a design-oriented lens. We present two interview studies conducted in Finland in 2023. The first study (n = 25) examines how people manage plastic waste at home, how they store it, where confusion arises, and which practical barriers interfere with sorting. The second study (n = 41) examines consumer views of laundry detergent packaging, a product category common to most households. Several alternative packaging formats have appeared on the market, yet little is known about how these formats influence recycling practices, transport to collection points, or perceptions of environmental impact. Focusing on this product group offers a concrete way to observe how packaging size, shape, and material interact with home routines.

Across both studies we identify recurring challenges related to space, ease of handling, clarity of instructions, and overall confidence in sorting choices. These themes point towards concrete opportunities for design interventions that support domestic routines and improve sorting quality. Building on these findings, we present four design concepts developed in a master's level student course. These concepts suggest practical ways to support home plastics recycling by addressing storage, guidance, and user experience within everyday domestic settings.

\section{Background}

This section outlines key ideas related to circular approaches to plastics and summarises the conditions that generally shape plastic packaging flows.

\subsection{Circular Approaches to Plastics}

Plastic products still commonly follow a linear pattern: materials are sourced, items are produced, used for a short period, and later discarded through incineration or landfill \cite{dijkstra2020business}. Circular approaches aim to reduce this pattern by promoting longer use, simplified material cycles, and improved recovery of resources \cite{johansen2022review,EC2018Plastics}. The literature describes several strategies for slowing or closing loops, ranging from reducing demand to repairing, reusing, and recycling products and materials \cite{potting2017circular}. Product design plays a central role in this shift, as structural choices, material combinations, and labelling influence durability, repair options, and recyclability \cite{Bocken2016,den2017product}.

For plastics, avoiding complex material mixes, enabling separation of components, and supporting clear communication about how items should be handled at end of life all contribute to improved recovery. Prior research points to quality problems when collected plastic waste is contaminated or contains mixed materials, which then limits how recycled material can be used \cite{eriksen2019quality}. Alongside industrial processes, everyday practices in homes strongly affect the outcomes. If sorting feels confusing or inconvenient, householders may avoid it or make incorrect choices, reducing both quality and quantity in collection streams \cite{oluwadipe2022critical}. This connection between design, domestic routines, and waste flows forms the foundation for our two studies.

\subsection{Plastic Packaging Flows in Finland}

In Finland, plastic packaging waste is managed through extended producer responsibility schemes. In 2019, approximately 76\,573\,t of post-consumer plastic packaging entered the market, with around 20\,430\,t collected through regional points and household systems \cite{judl2023towards}. The remainder was disposed of in mixed municipal waste. From the separately collected fraction, roughly 18\,000\,t was mechanically recycled, while the reject stream was directed to energy recovery. Recycled output consists mainly of LDPE, HDPE, PP, and PET.

%The effectiveness of recycling is influenced by both the quantity of material and its quality. Contamination and heterogeneity continue to restrict the share of collected plastics that can be processed into new products \cite{eriksen2019quality}. Household sorting plays a key part in this process,  individual decisions about storage, cleaning, and correct identification of plastics feed directly into the wider system.

These conditions set the context for the two studies presented here, which draw on a Finnish participant sample to examine domestic plastic handling and how packaging characteristics influence recycling practices.

\section{Method}

Two qualitative studies were carried out to examine how people handle plastic waste in their homes and how the form of plastic packaging shapes their recycling habits. Study~1 focused on everyday sorting and storage in domestic settings. Study~2 examined laundry detergent packaging as a concrete case of large plastic containers that often require extra handling effort.

\subsection{Study 1: Domestic Plastic Sorting Practices}

Study~1 used short, semi-structured interviews to explore how people organise, store, and dispose of plastic packaging at home. The aim was to capture the points in their routines where uncertainty, inconvenience, or practical obstacles occur. Discussions covered storage solutions, transport to collection points, situations where plastic is placed in mixed waste, and ideas participants felt could improve sorting.

Participants were recruited through convenience sampling in everyday social settings. A total of 25 people took part. Interviews lasted approximately ten minutes and were recorded as written notes and short transcripts. Before each interview, participants were informed that responses would remain anonymous and that the study sought to understand general challenges rather than evaluate individual behaviour. Consent was obtained from each participant.

Analysis followed an inductive approach. Transcripts were read multiple times, coded for recurring themes, and grouped into broader categories describing storage challenges, uncertainties about correct sorting, difficulties with specific items, and ideas for improvement.

\subsection{Study 2: Laundry Detergent Packaging and Recycling}

\begin{figure}
    \centering
    \includegraphics[width=0.75\linewidth]{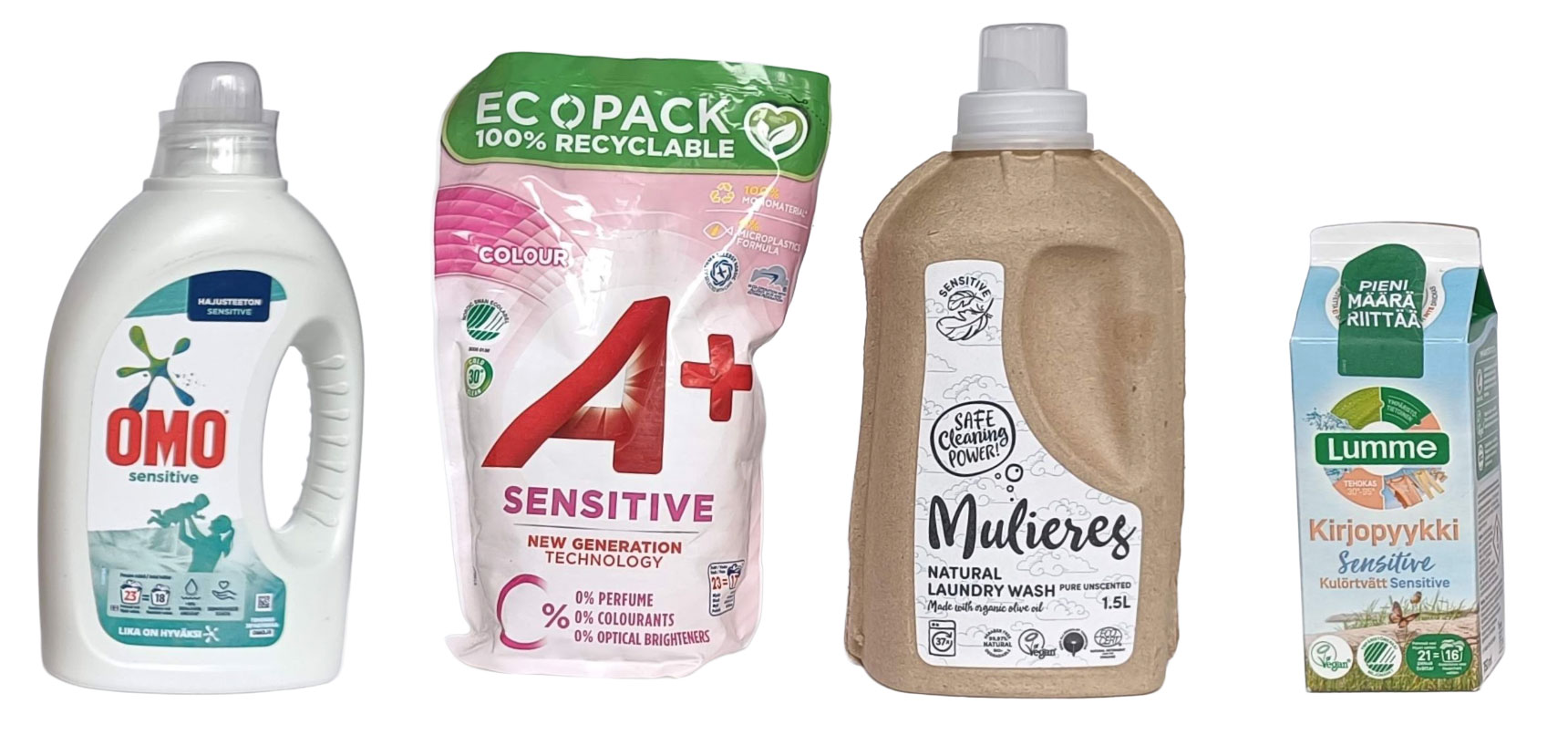}
    \caption{The four laundry detergent packages used as probes in the interviews.}
    \label{fig:washing_liquid}
\end{figure}

Study~2 examined a specific product category that commonly appears in household plastic waste streams: liquid laundry detergent containers. The study combined brief interviews with a structured questionnaire. The goal was to understand how the size, material, and shape of detergent containers influence purchasing decisions, handling in the home, and motivation to recycle. A set of four physical laundry detergent packages (Figure~\ref{fig:washing_liquid}) served as discussion prompts. Handling the containers helped participants articulate what feels practical or impractical in daily routines, such as weight, shape, stability, ease of emptying, and how packages fit into home storage or transport to collection points.

Data were collected in Rovaniemi, Finland, in July 2023. Most participants were approached in public spaces such as pedestrian routes and parks, while a smaller group responded to a call posted in three local hobby-related WhatsApp groups. In total, 41 people took part. Conversations were conducted face to face and followed a fixed set of topics covering purchase habits, experiences with detergent containers, views on recyclability, and practical obstacles in sorting.

%Demographic information was collected to give context to responses. 
Of the 41 participants, 26 identified as women and 13 as men, while two chose not to report gender. Ages ranged from young adults to senior citizens, with the largest group between 31 and 45 years old. Open responses were analysed thematically. Notes and transcriptions were coded to identify patterns related to purchase motivations, packaging preferences, obstacles in recycling, and areas where people felt improvements would make sorting easier.

\section{Results}

The two studies reveal consistent patterns in how people manage plastic waste at home and how detergent packaging shapes everyday routines. Study~1 highlights frictions in storage, sorting, and transport. Study~2 offers a more detailed look at how one common household product group generates practical challenges and influences recycling behaviour.

\subsection{Study 1: Domestic Sorting Challenges}

The interviews report that storage for plastic waste in the home is a central concern. Many participants described improvised solutions such as bags hung on cupboard doors, buckets under the sink, or interim piles on countertops. These arrangements were often described as temporary, visually messy, or difficult to keep clean. Several participants noted that the volume of plastic packaging grows faster than other materials, which creates pressure to empty storage more frequently.

Uncertainty about correct sorting was another recurring theme. Participants reported confusion about whether certain materials belong in plastic recycling, especially multi-layer films, rigid items such as toys, and packaging with mixed components. Some participants said they choose mixed waste if they feel unsure, either to avoid contaminating the plastic stream or due to limited time to check instructions.

Practical obstacles with specific items were also prominent. Bulky containers, greasy food packaging, and items that require thorough rinsing were frequently mentioned. Some participants said they avoid rinsing with hot water if they feel it wastes energy, while others noted that cold rinsing is not always effective. This occasionally led to packaging being discarded with mixed waste. Access to collection points influenced how often plastic is taken out of the home. Participants with a nearby property-level bin described stable routines, whereas those who needed to travel to city drop-off stations said that distance or weather sometimes delayed disposal. 

Participants offered a range of ideas for improvement, including clearer labelling, easier-to-clean packaging, and home storage solutions that reduce clutter. Many of these ideas relate directly to design opportunities explored later in the paper.

\subsection{Study 2: Views on Laundry Detergent Packaging}

Study~2 provides a closer look at a single product group that, in this Finnish sample, is mainly encountered as large plastic containers. Participants compared four physical packaging options and described how these formats fit into their routines. When discussing their current detergent purchases, participants emphasised price, familiar brands, scent, and the expectation that the product cleans well. Environmental considerations were mentioned, but typically after the practical factors. Several participants said they try to choose packaging that seems easier to recycle, yet added that they seldom check detailed information when shopping.

Handling the physical packages brought up concerns about weight, rigidity, ease of pouring, and how well the containers empty. Larger bottles were described as heavy, awkward to hold, or slow to drain fully. Flexible pouches were sometimes seen as more compact but also harder to clean or sort due to uncertainty about the materials. Some participants expressed doubts about the recyclability of certain composite pouches. Participants also discussed what happens to the containers at home. Large bottles were often stored in utility rooms or bathroom shelves, and some mentioned that the size creates clutter under sinks or in laundry areas. A few participants said they sometimes place detergent containers in mixed waste if they feel unsure about rinsing or if they find the container difficult to handle.

Several respondents spoke positively about the idea of refillable or concentrated formats, but noted that these options are not widely available in their local shops. Others pointed out that refill systems require space and clear instructions to feel practical in daily life. Across the sample, the most frequently mentioned challenges were:  
(1) uncertainty about materials,  
(2) difficulty rinsing or fully emptying containers, and  
(3) limited space for temporary storage before disposal.  These findings align closely with the themes in Study~1 and highlight a shared set of pressures that shape domestic recycling practices.

\section{Design Concepts}

To extend the insights from the two studies, a set of design concepts was created by students in the master's level design course at the University of Lapland, Finland, in autumn~2023. The assignment formed one of three group projects on the course and was linked with the PlastLIFE project, which acted as the client for the work. PlastLIFE focuses on reducing the use of plastics, replacing problematic materials, and promoting recycling, which provides a clear framing for exploring domestic plastic handling.

Altogether eighteen students took part, organised into four groups of four to five members. The participants represented a mix of backgrounds, including design, audiovisual media, sustainable art and design, and international exchange students in design. %Teaching was carried out in English, and the course included both Finnish and international students.

\begin{figure}
    \centering
    \includegraphics[width=1\linewidth]{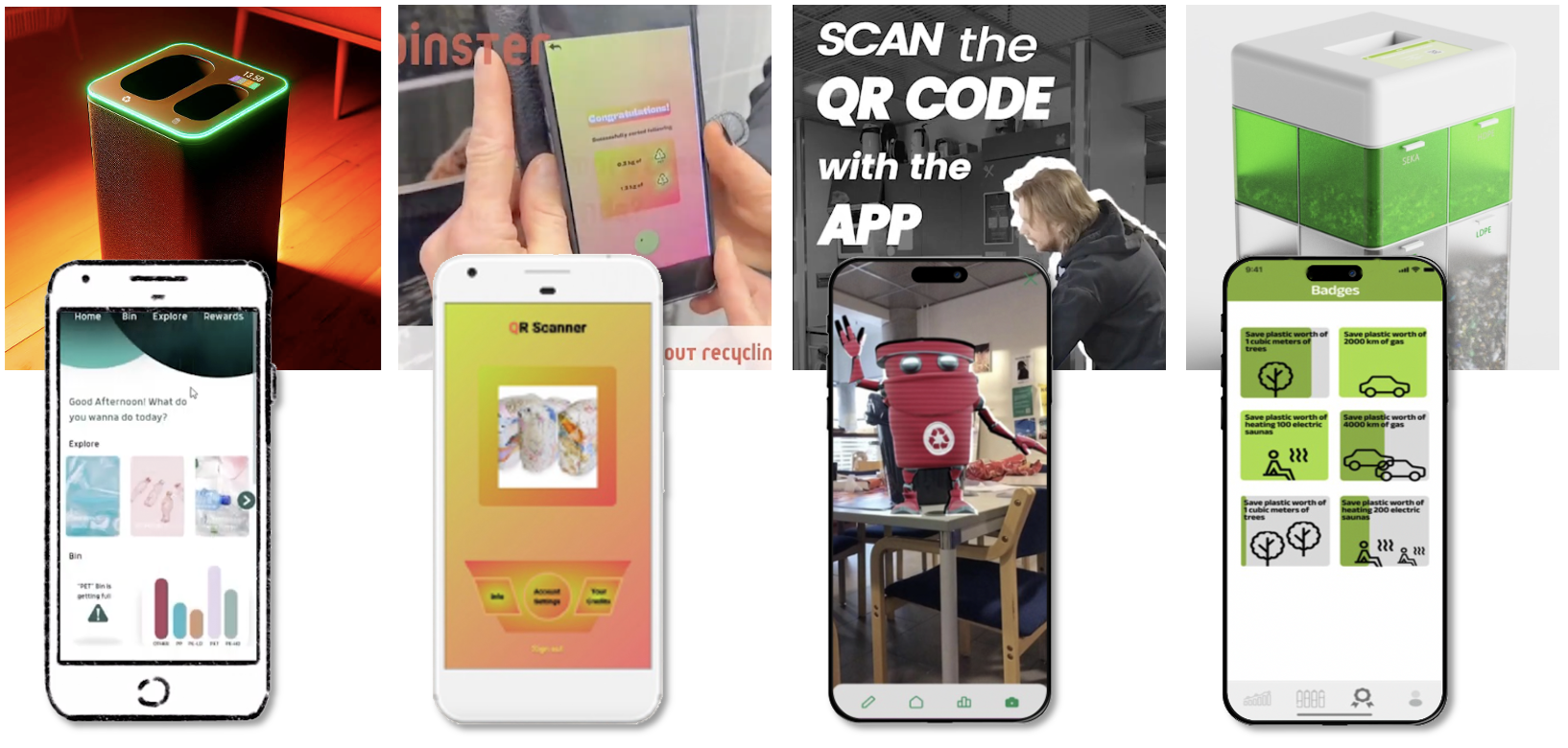}
    \caption{The four concepts. Left to right: Plasticle, Binster, Plasmate, PolyBin.}
    \label{fig:four_concepts}
\end{figure}

The task brief asked each group to create an interactive bin for plastic waste and a connected mobile application. Students were encouraged to define a clear goal for their concept, such as supporting better sorting routines, reducing clutter at home, improving recognition of packaging types, or helping households reflect on how much plastic they produce. Each group delivered a presentation, design materials, and a short demonstration video. The resulting concepts form four distinct approaches to supporting domestic plastic recycling (Figure \ref{fig:four_concepts}). The following subsections outline the four design concepts.

\subsection*{Plasticle}

Plasticle is an interactive recycling system that integrates a sensor-equipped bin with a mobile application to support residents who are poor recyclers, e.g., due to limited space or low motivation. The bin has two openings that lead to internal containers equipped with sensors and a crushing mechanism. These components provide automated waste compaction, light-based feedback, and a small screen interface, while colour coding communicates different plastic categories at a glance.

The accompanying application connects to the bin, shows fill levels, statistics on the amount and types of plastics collected, and provides information, news, and notifications related to plastic recycling. Users accumulate tokens through regular use that can be exchanged for money-off shopping coupons, linking pro-environmental behaviour with tangible rewards and positioning Plasticle as both an informational and incentive-based intervention in the home.

\subsection*{Binster}

Binster is a family-oriented interactive recycling concept that combines a talking sensor-based bin with a reward-focused smartphone application, with the specific goal of forming recycling habits in children and youth. The bin incorporates scanners to recognise plastic types, internal mechanisms to sort them into appropriate compartments, and an expressive character-like exterior that reacts when it receives ``food'' in the form of plastic waste.

Once full, the system signals through sound and triggers a notification in the application, which tracks how much each family member has recycled, awards credits for every contribution, and allows these credits to be converted into money or prizes. The interface includes QR-based linkage to the bin, account management, and progressive bonus levels. This supports friendly competition inside the household around who recycles the most and frames plastic sorting as a playful, rewarded practice in everyday life.

\subsection*{Plasmate}

Plasmate introduces a low-threshold smart bin concept in which the only physical component is a recyclable plastic bag with a QR code that can be fitted into any existing kitchen bin. This is combined with an augmented reality application that animates a virtual trash bin character. The concept targets families with children in single-family homes and frames plastic recycling as a shared activity in which children feed clean plastic items to the mascot, gain points, and use these points to customise its appearance.

Through Augmented Reality (AR) views in the mobile app, the character appears in the home environment and reacts to user actions. The application tracks contributions, visualises weekly and monthly recycling statistics, and supports neighbourhood, city, and national competitions through leaderboards. By decoupling the service from a dedicated hardware bin and emphasising playful social comparison, Plasmate aims to increase engagement with plastic recycling without major changes to kitchen infrastructure.

\subsection*{PolyBin}

PolyBin is a smart household recycling station that combines a modular trash bin with a companion mobile application to support sorting of different plastic types in everyday family life. The physical bin consists of stackable containers of varying sizes that can be configured to fit individual households. Transparent lower modules display the accumulated shredded plastic, and colour-coded labels highlight categories such as PET, HDPE, PVC, LDPE, PP, PS, and mixed waste.

A screen on the top unit guides users, shows error messages when an item is detected as non-plastic, and prompts users to connect the bin to the app through a QR code. The mobile interface presents weekly and annual statistics, feedback on sorting performance, and educational content about plastic types. Routine disposal activity is transformed into a data-rich practice that makes the volume and composition of household plastic waste visible over time.

\subsection*{Concept comparison}

\begin{figure}
    \centering
    \includegraphics[width=0.5\linewidth]{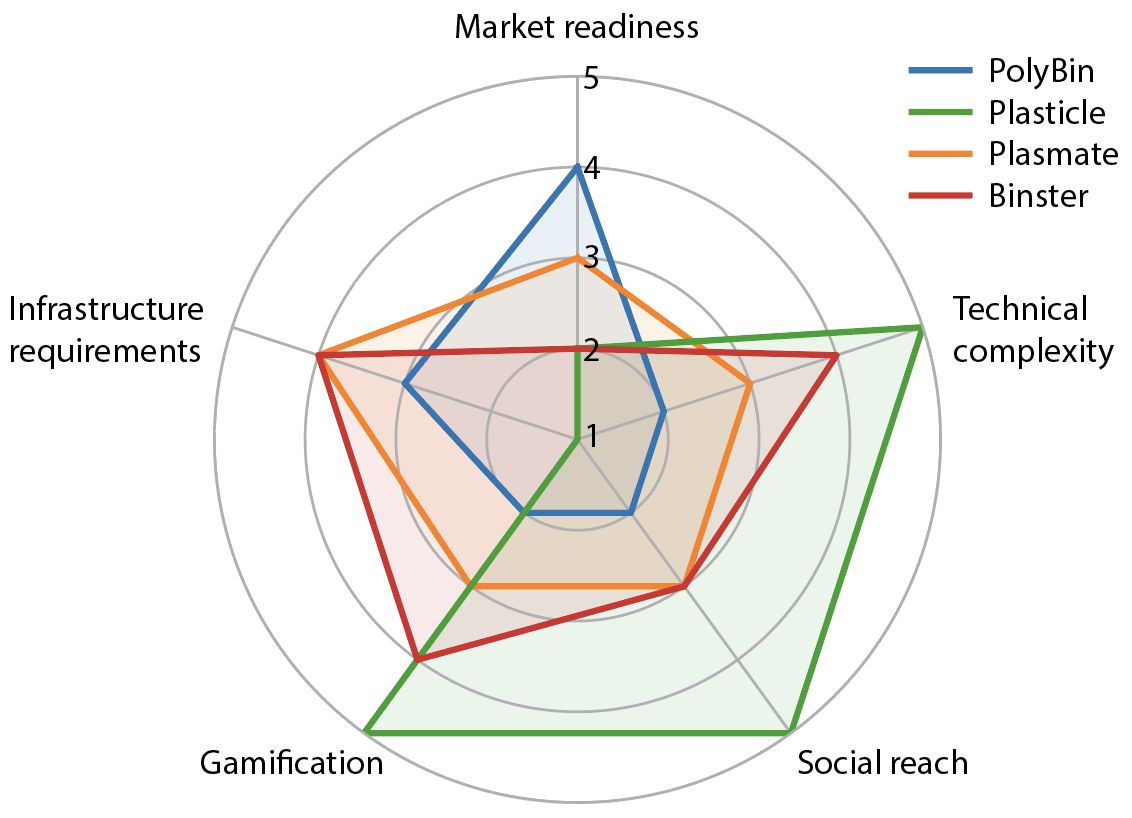}
    \caption{Evaluated characteristics of the four home recycling concepts.}
    \label{fig:concept_radar}
\end{figure}

To compare the four concepts, two experienced design teachers familiar with interaction design and sustainable HCI rated each concept independently. The rating used five criteria that reflect themes raised in the studies and in the design reflection: (1) market readiness, (2) technical complexity, (3) social reach, (4) gamification and play, and (5) level of disruption to existing infrastructure. Each criterion was rated on a scale from 1 (low) to 5 (high). The criteria were selected to capture both feasibility and scope: market readiness and infrastructure disruption describe how easily a concept could be integrated into current systems, technical complexity describes the demands of sensing, automation, and back-end services, and social reach and gamification describe the intended range of users and the strength of playful elements. Figure~\ref{fig:concept_radar} presents the average ratings across the two evaluators as a radar chart.

\section{Discussion}

The two empirical studies highlight challenges in the domestic phase of plastic recycling and point to opportunities for design interventions at the level of products, home infrastructure, and digital services. The four design concepts illustrate how such interventions might be shaped in practice, but they also open questions about feasibility, long-term engagement, and the ecological impact of new devices.

Our findings confirm that the simplicity of the sorting process strongly shapes household recycling behaviour, in line with earlier work \cite{kokkonen2020kierratyksen}. Participants expressed a clear preference for straightforward instructions and easily recognisable categories. Confusion about multi-material items, films, and certain rigid plastics led some respondents to discard items in mixed waste to avoid perceived contamination of the plastic stream. This echoes concerns raised about quality losses in plastic recycling due to contamination and heterogeneity \cite{eriksen2019quality}.

The Finnish context amplifies these issues. Plastic packaging recycling rates are not yet aligned with upcoming EU targets, and the separation stage in homes remains a weak point in the chain. The laundry detergent case shows how large, heavy containers, mixed-material pouches, and uncertainty about rinsing combine to create friction. These factors affect both the quantity and the quality of material reaching collection systems, reinforcing the need for better product design and clear labelling, alongside consistent guidance from waste management organisations and public agencies \cite{EEA2024,RecyclingPartnership2021}.

Participants in both studies described a lack of storage space as a constant pressure, consistent with prior research on household recycling barriers related to inconvenience and limited storage options \cite{oluwadipe2022critical}. Plastic waste was often stored in improvised containers or bags that quickly filled up and created clutter. Many respondents expressed interest in more compact packaging and better-fitting bins rather than more complex routines. This aligns with actions higher in waste hierarchies focused on source reduction and packaging minimisation \cite{potting2017circular}, and with design research showing that everyday domestic infrastructure and routines shape sustainable behaviours \cite{Comber2013Designing}, but still leaves a need for solutions that make the remaining sorting and storage more manageable.

Product design influenced purchasing decisions in nuanced ways. Price, scent, brand familiarity, and perceived cleaning performance were primary drivers in detergent choices, consistent with research showing that practical considerations often outweigh environmental criteria in household purchasing decisions \cite{Young2010Green,Joshi2015GreenPurchase}. Many participants nonetheless expressed concern for environmental impact, and packaging that seemed easier to recycle or that signalled environmental benefits was viewed positively, in line with evidence that visual and verbal sustainability cues shape consumer responses \cite{Magnier2015EcoPackaging}. This pattern reflects the well-documented value–action gap \cite{Kollmuss2002Gap}, in which stated environmental concerns do not always translate into final choices. Taken together, the results suggest that product and packaging design should support both functional needs and environmental values, e.g., through clear labelling, container forms that are easy to empty, and materials compatible with local recycling systems.

\subsection*{Design concepts as exploratory responses}

The four student concepts extend these findings into concrete design directions. PolyBin and Plasticle introduce dedicated bin systems with modular compartments, sensors, and compaction mechanisms. They respond directly to the reported space constraints and the need for clearer guidance on plastic types. Transparent modules, colour coding, and built-in feedback make plastic flows more visible in the home and connect sorting actions to statistics over time. Similar smart-bin initiatives have shown the value of combining sensing, user identification, and norm- or emotion-based feedback in waste systems \cite{Guna2022SmartBin}.

Plasmate and Binster take a more speculative stance. Plasmate relies on QR-coded bags and an augmented reality mascot, bringing recycling into children’s play and social comparison between households. Binster combines automatic recognition and sorting of plastics with a character-like bin and reward schemes that track contributions at the level of individual family members. Both concepts use gamification, anthropomorphism, and competition to support habit formation. This design space is closely related to evaluations of emoticon-based recycling bins and smart recycling systems, where playful feedback, points, and rewards have been shown to increase recycling behaviour and intention to use such systems \cite{Berengueres2013Emobin,Liu2022SmartRecycling}.

Across all four concepts, a shared pattern emerges: each pairs a physical artefact (bin, bag, or container system) with a digital layer that offers feedback, statistics, and educational content. The concepts address low motivation, limited knowledge of plastic categories, and weak visibility of waste volumes by making plastic more present in daily routines through visualisations and messages at the moment of disposal. Habit formation is treated as a central concern, using points, levels, comparisons, or rewards to keep residents engaged. Prior work on anthropomorphic conversational agents for sustainability suggests that personified agents can strengthen users’ sense of connection and support eco-related self-efficacy, even when their effect on measured outcomes is mixed \cite{Giudici2025AnthroCA}. This provides a useful backdrop for the character-based interactions proposed in Plasmate and Binster.

\subsection*{Feasibility and broader concerns}

The concepts span a spectrum from near-term interventions to more speculative visions. PolyBin and Plasticle sit closer to current practice, as they build on familiar bin formats and established sensing and display technologies. Plasmate and Binster demand more advanced recognition, AR, and large-scale infrastructures for competitions or monetary rewards. All four depend on alignment with local waste management guidelines and infrastructures, and would require collaboration with municipalities and service providers before any real deployment.

At the same time, the concepts raise important questions. Long-term engagement is uncertain: playful characters and points may attract attention initially, but novelty may fade over time, as also suggested in broader work on gamified sustainability technologies \cite{Liu2022SmartRecycling}. Tracking waste volumes and individual contributions in households brings privacy and data governance concerns, especially if data leave the home or link to external reward systems. These concerns resonate with research on smart-home devices, where residents often feel uncertain about data flows, sharing practices, and ownership of the data collected in domestic environments \cite{Tabassum2019SmartHomeData}. Incentive schemes may carry equity risks if some households cannot participate due to connectivity, housing conditions, or time constraints. Finally, any new hardware for sustainability purposes must be assessed against its own material footprint; adding electronics and plastics for sorting devices can conflict with the aim of reducing resource use if not carefully designed and maintained \cite{Charfeddine2023ICTSustainability}.

%One way to capture these trade-offs is to compare the concepts along shared dimensions such as market readiness, technical complexity, social reach, degree of playfulness, and level of disruption to existing infrastructure. Figure~\ref{fig:concept_radar} summarises these ratings and makes visible how PolyBin and Plasticle sit closer to current practice, whereas Plasmate and Binster reach for higher technical complexity, stronger gamification, and wider social reach.

\subsection*{Limitations and future work}

The studies and concepts presented here have several limitations. Convenience sampling in one Finnish city means that the participant group may not represent national patterns, and there is a risk of social desirability bias, as participants might overstate their recycling activity in face-to-face interviews \cite{Grimm2010SocialDesirability}. The student concepts are early-stage designs created within a course context and have not been evaluated in real homes. Their feasibility and impact remain untested.

Future research could combine qualitative studies with observational data or measured waste flows to gain a clearer picture of actual behaviour. Prototyping one or more of the concepts in collaboration with municipalities or housing companies would allow testing of how interactive bins and apps fit into everyday routines, how residents respond to feedback and incentives, and what kinds of support are needed from packaging designers, retailers, and waste management actors. Experiences from situated energy-feedback projects such as the Tidy Street initiative in Brighton, UK, show that visible, context-specific feedback can influence household behaviour \cite{TidyStreet2011}, suggesting that similar approaches could be explored in recycling contexts. Any such work should track environmental effects carefully, so that interventions aimed at improving recycling genuinely support a shift to more sustainable plastic use.

\section{Conclusion}

This article examined how people in Finland handle plastic packaging in their homes and how the design of everyday products shapes recycling practices. Two empirical studies highlighted recurring challenges in the domestic stage of recycling, including limited storage space, uncertainty about sorting rules, and the effort required to clean and handle large containers. Participants valued clarity, simplicity, and packaging that fits smoothly into existing routines, yet many expressed interest in reducing plastic use more broadly. These findings point to a continued need for packaging that is easier to empty, easier to recognise, and better aligned with local recycling systems.

The four design concepts developed by master’s-level design students illustrate different ways interactive bins and digital services might support households. They use visual feedback, playful engagement, data displays, and automation to address low motivation, limited knowledge, and lack of visibility of plastic waste. Their diversity also surfaces important considerations for future work, including long-term engagement, data practices in domestic spaces, and the environmental footprint of new hardware.

\begin{acks}
This study was funded by the Rethinking Plastics in a Sustainable Circular Economy (PlastLIFE) project (LIFE21-IPE-FI-PlastLIFE). The PlastLIFE project was co-funded by the European Union. Views and opinions expressed are however those of the authors only and do not necessarily reflect those of the European Union or CINEA. Neither the European Union nor the granting authority can be held responsible for them.
\end{acks}

\bibliographystyle{ACM-Reference-Format}
\bibliography{sample-base}

\end{document}